\documentclass[11pt,a4paper]{article}
\usepackage{jheppub}

\usepackage{amsmath, amsfonts, amsthm, amssymb, graphicx}
\usepackage[dvipsnames]{xcolor}
\usepackage{epstopdf}
 \usepackage{slashed}
\usepackage{changepage}
 \usepackage{vector}
\usepackage{soul}

\usepackage{pifont}

\usepackage{caption}
\def\be{\begin{equation}}
\def\ee{\end{equation}}
\def\ba{\begin{eqnarray}}
\def\ea{\end{eqnarray}}

\def\bdm{\begin{displaymath}}
\def\edm{\end{displaymath}}

\def\bq{\begin{quote}}
\def\eq{\end{quote}}

\def\del{\partial}
\def\ltap{\ \raise.3ex\hbox{$<$\kern-.75em\lower1ex\hbox{$\sim$}}\ }
\def\gtap{\ \raise.3ex\hbox{$>$\kern-.75em\lower1ex\hbox{$\sim$}}\ }
\def\gl{\ \raise.5ex\hbox{$>$}\kern-.8em\lower.5ex\hbox{$<$}\ }
\def\roughly#1{\raise.3ex\hbox{$#1$\kern-.75em\lower1ex\hbox{$\sim$}}}

 at 10truept
\def\muv{M_\text{UV}}

\newcommand{\beq}{\begin{equation}}
\newcommand{\eeq}{\end{equation}}
\newcommand{\bea}{\begin{eqnarray}}
\newcommand{\eea}{\end{eqnarray}}
\newcommand{\beqa}{\begin{eqnarray}}
\newcommand{\eeqa}{\end{eqnarray}}

\def \del {\partial}

\def \vqft {V_\text{QFT}}

\def \is {{i}}

\newcommand{\order}{{\cal O}}

\def\be{\begin{equation}}
\def\ee{\end{equation}}

\usepackage[dvipsnames]{xcolor}

\title{The cosmological constant and the weak gravity conjecture}

\author[1,2]{Yang Liu,} 
\author[1,2]{Antonio Padilla} 
\author[3,4]{and Francisco G. Pedro}

 \affiliation[1]{School of Physics and Astronomy, University of Nottingham, University Park, Nottingham NG7 2RD, United Kingdom}
\affiliation[2]{Nottingham Centre of Gravity, University of Nottingham, Nottingham NG7 2RD, UK}
 \affiliation[3]{Dipartimento di Fisica e Astronomia, Universit\`a di Bologna, via Irnerio 46, 40126 Bologna, Italy}
\affiliation[4]{INFN, Sezione di Bologna, viale Berti Pichat 6/2, 40127 Bologna, Italy}

\emailAdd{yang.liu@nottingham.ac.uk}
\emailAdd{antonio.padilla@nottingham.ac.uk}
\emailAdd{francisco.soares@unibo.it}

\abstract{We examine the descent via membrane nucleation through a landscape of vacua where the cosmological constant is given by a combination of four-form fluxes. It has been shown that this descent can be slowed exponentially for very low curvature vacua close to Minkowski space in a wide class of models satisfying certain parametric conditions, providing a possible solution to the cosmological constant problem.   We explore in detail whether or not those parametric conditions are compatible with the membrane weak gravity conjecture. Whilst it is true that there is often tension, we show that this is not always the case and present an explicit model where Minkowski space is absolutely stable and the weak gravity conjecture is satisfied. This corresponds to an extension of the Bousso-Polchinski model into a generalised DBI action for four-forms.  We also clarify how the landscape should be populated in a consistent model.
} 

\begin{document}
\maketitle
\section{Introduction}
In contrast to our planet, our universe is a surprisingly quiet and gentle place, its weakly accelerated expansion driven by a dark energy whose density is at least thirty orders of magnitude less than that of water.  This dark energy is consistent with a vacuum energy, or cosmological constant, whose tiny density remains constant over time. However, when we estimate the contribution to the vacuum energy from quantum fields, we find that its natural value scales like the fourth power of the field theory cut-off.  Taking that cut-off to lie at  the TeV scale or beyond, the energy density of the vacuum is then expected to lie at least sixty orders of magnitude higher than the scale of dark energy.  This is a problem known as the cosmological constant problem. For reviews, see \cite{Weinberg:1988cp,Polchinski:2006gy,Burgess:2013ara,Padilla:2015aaa,Bernardo:2022cck}.

In a universe with a landscape of different vacua, as one might expect to occur in string theory,  we can imagine a scenario in which the cosmological constant starts out large but descends to smaller values via some process, until it reaches the small value we see today. Brown and Teitelboim proposed an interesting toy model in which a landscape of different vacua is provided by the vacuum expectation value of a four-form field strength and the vacua are scanned by the nucleation of membranes charged under the corresponding three-form field \cite{Brown:1987dd,Brown:1988kg}.  This works because the four-form field strength  gravitates like a cosmological constant in four spacetime dimensions, its energy density correcting the renormalised vacuum energy by an amount proportional to the square of the four-form flux.  In the presence of membranes, the flux is quantised in units of the membrane charge, changing by a single unit whenever a membrane is crossed.  As a result, the overall cosmological constant changes by a discrete amount whenever a membrane is nucleated through quantum effects, allowing the cosmological constant to descend  from large to small values. 

The problem with the Brown and Teitelboim proposal  is that the descent is too slow. In order for the landscape to be sufficiently dense to readily accommodate vacua with energy densities around the dark energy scale, we must assume that the membrane charge is extremely small in units of a Planckian cut-off. Descent from high scale vacua with energy densities close to the cut off to the low density vacuum we see today proceeds via a large number of intermediate steps, slowly descending from one cosmological constant to another.  The time scales are such that the universe undergoes long periods of accelerated expansion, diluting away all matter and leaving us with a cold and empty universe. The empty universe problem also spoils a related mechanism  for slowly descending through adjacent vacua in a ``washboard" potential for a scalar field, originally proposed by Abbott \cite{Abbott:1984qf}. 

Bousso and Polchinski showed how the empty universe problem could be avoided by extending the Brown and Teitelboim mechanism to a large number of four-forms \cite{Bousso:2000xa}. This allows for a sufficiently dense landscape without fine tuning the membrane charges to very small values. The latter control the change in the cosmological constant whenever we have membrane nucleation and, as they are just an order or magnitude or so below the cut-off, it is possible to descend from a high scale vacuum to the one we see today in a single jump. The empty universe problem can then be avoided as follows.  In the moments before the jump, the underlying vacuum energy is large and the inflaton field is displaced to large values thanks to quantum diffusion. Immediately after the jump, the minimum of the inflaton potential and the underlying vacuum energy can be small but the inflaton itself is still displaced. The universe undergoes a period of slow roll inflation until the inflaton rolls down its potential and begins to oscillate about the minimum, allowing the universe to reheat. The empty universe problem can be avoided in a similar way with just two species of four-form by assuming that the corresponding membranes have an (almost) irrational charge ratio \cite{Kaloper:2022oqv,Kaloper:2022yiw,Kaloper:2022jpv}.  Note that Garriga and Vilenkin have argued that  the empty universe problem is a fallacy, at least in the context of eternal inflation \cite{Garriga:2000cv}. 

This type of descent from high to low scale vacua is able to explain how we can arrive at the current vacuum, but it doesn't explain why. Bousso and Polchinski argue that the current low curvature vacuum is selected on anthropic grounds, although the validity of this claim has been criticised in \cite{Banks:2000pj}. Alternatively, we might look for ways to slow down the descent, or even stop it altogether, whenever the curvature approaches zero. In \cite{Feng:2000if}, the authors consider the Brown-Teitelboim model, but take into account  membrane stacks and degeneracy factors. They claim that the tunnelling rates depend on the ambient de Sitter temperature, being enhanced at high scales and switching off at low scales. However, Garriga and Vilenkin argue the opposite: that these enhancement factors are independent of the ambient de Sitter temperature and do not switch off at low curvature \cite{Garriga:2000cv}.

A mechanism for halting the descent is also presented in \cite{Kaloper:2022oqv,Kaloper:2022yiw,Kaloper:2022jpv}. Indeed, in \cite{Kaloper:2022jpv}, the authors consider a generalisation of Henneaux and Teitelboim's covariant formulation of unimodular gravity \cite{Henneaux:1989zc} with two species of four-form.  By including charged membranes with an (almost) irrational charge ratio, they obtain a dense landscape of vacua that includes the current vacuum. Transitions between vacua go through via membrane nucleation. By making some relatively mild assumptions on the membrane charge and tensions, the nucleation rate  slows down exponentially quickly, halting the decay of the vacuum at vanishing curvature. 

In \cite{us}, this mechanism was shown to extend to a much larger class of models, including the classic framework of Bousso and Polchinski.  This meant there was a way to halt the descent between vacua at vanishingly low curvature in a wide class of effective field theories, providing an interesting alternative to anthropic selection.  The family of theories included multiple species of four-forms coupled to multiple scalars with generic potentials, increasing the hope that the halting mechanism could emerge naturally as a low energy limit of string compactifications. However, in this work it was also noted that the corresponding condition on the membrane charges and tensions was at odds with the membrane version of the weak gravity conjecture (WGC) \cite{Arkani-Hamed:2006emk,Ibanez:2015fcv}, especially in the case of the Bousso-Polchinski model.

In this paper, we further explore the dynamics of this wide class of models, paying close attention to descent between vacua and the role played by  the WGC.  A related analysis was recently  performed in \cite{Kaloper:2023vrs}, although our more general approach allows us to clarify some important points. In particular, there is certainly a tension between the WGC and the halting conditions for a canonical theory  quadratic in the flux, as already pointed out in \cite{us}. However, if a higher power of the flux dominates the dynamics, these tensions can be eased by assuming a  hierarchy between the renormalised vacuum energy and the cut-off scale, as one  could easily obtain in the presence of supersymmetry. Of course, the dominance of a single higher power is not consistent with a well defined effective theory at weak coupling. This motivates us to study the dynamics at strong coupling, taking into account a full tower of operators. We first consider the generic case using naive dimensional analysis \cite{Manohar:1983md,Gavela:2016bzc} and once again find tension between the WGC and the halting conditions. However, we also explore an elegant extension of the Bousso-Polchinski model based on a generalised form of the DBI action \cite{Jurco:2012yv, Ho:2014una}.  This has the property that it behaves like a Bousso-Polchinski set-up at smaller values of the four-form flux, although at large values it behaves more like the linear four-form proposal presented in \cite{Kaloper:2022jpv}.  As we will  show explicitly, this generalised model can be taken to be compatible with the WGC {\it without} spoiling the halting of vacuum descent at very low curvatures.

Our analysis of the dynamics also helps us to better understand how very low curvature vacua are populated. In particular, following  \cite{Kaloper:2023vrs}, we might consider two distinct phases in the membrane nucleation rate: the boiling phase, where the rate is high and the braking phase, where the rate is low. Having arrived at a very low curvature vacuum  with further descent almost halted, we can ask how we got here. Did the universe start out in a high curvature vacuum and boil rapidly down to the current low curvature vacuum? Or did we reach the current vacuum having already entered the braking phase before the last transition?  Under a very mild set of assumptions, we will show that it must be the latter.  This now suggests the following set-up: the universe starts out in a high curvature vacuum and boils down to a bubbling soup of vacua of intermediate curvature. Within those bubbles of intermediate curvature we are now in the braking phase and future transitions are slow. Nevertheless, given enough time, we do reach the low curvature vacua like ours, which are even more stable than their ancestors.  These low curvature vacua are by far the longest lived. 

The rest of this paper is organised as follows.  In section \ref{sec:review}, we review the main results of \cite{us} applied to an effective theory of four-forms, computing  transition rates between vacua and conditions found for bubble nucleation to come to a halt once we reach vanishing vacuum curvature. In section \ref{sec:models}, we study specific models, including those for which the cosmological constant is a homogeneous polynomial of the four-form flux and the generalised DBI set-up, with a view to understanding the constraints imposed by both the WGC and the halting condition.  In section \ref{sec:dsdecay} we consider the dynamics of vacuum decay in a general setting focussing on parametric behaviour of the rate of decay.  In section \ref{sec:discuss},  we discuss our findings and summarize some  remaining concerns.

\section{Membrane nucleation in effective theories of four-forms}\label{sec:review}
We begin with a generalised theory of four-forms on a manifold, $\mathcal{M}$, with a dynamical metric $g_{\mu\nu}$ described by the following action, 
\be\label{Lorentzianaction}
S =\int_\mathcal{M} d^4 x\sqrt{|g|} \left[\frac{M_{pl}^2}{2}  R+f(\star F_i) \right]
 +S_\text{boundary}+S_\text{membranes}, 
\ee
where we have a family of three-form fields, $A^i=\frac{1}{3!} A^i_{\mu\nu \alpha} dx^\mu \wedge dx^\nu \wedge dx^\alpha$, with corresponding field strength $F^i=dA^i$ and  $\star$ denotes the Hodge star operator on the manifold.  $R$ is the Ricci  scalar on the manifold while the function $f$ is assumed to admit an expansion of the form 
\be \label{f}
f(\star F_i)=\sum_{n\geq 2} \frac{a_{i_1 \ldots i_n}}{ n! \muv^{2n-4}} (\star F^{i_1}) \ldots  (\star F^{i_n}),  
\ee
where $\muv\lesssim M_{pl} $ is the cut-off of the EFT and  the dimensionless coefficients $a_{i_1 \ldots i_n}$ are naturally $\order(1)$. 

The boundary,  $\partial \mathcal{M}$, is taken to be a co-dimension one surface described by the embedding $x^\mu=X^\mu(\xi^a)$, such that $\gamma_{ab}=g_{\mu\nu}\del_a X^\mu \del_b X^\nu$ is the induced metric on the boundary and $\alpha^i=\frac{1}{3!} A^i_{\mu\nu \alpha} \del_a X^\mu \del_b X^\nu \del_c X^\alpha d\xi^a \wedge d \xi^b \wedge d\xi^c $ is the pull back of the three-form, $A^i$. We will describe the boundary action explicitly in a moment, after switching to a dual formulation.  We also include contributions from membranes and anti-membranes, $\Sigma_I$,  charged under any of the three-forms, such that
\be \label{membact}
S_\text{membranes}=-\sum_I \left\{\eta^i_I q_i\ \int_{\Sigma_I} \alpha_I^i + |\eta^i_I| \tau_i \int_{\Sigma_I} d^3 \xi \sqrt{|\gamma_I}| \right\},
\ee
where  $\gamma^I_{ab}$ is the induced metric on $\Sigma_I$ and $\alpha_I^i$ is the pull-back of the three-forms. Membranes charged under $A^i$ carry a fundamental charge $\pm q_i$ depending on whether they are branes or antibranes and tension $\tau_i$. In the action \eqref{membact},  $\eta^i_I=0, \pm 1$ depending on whether the membrane $\Sigma_I$ carries positive ($\eta^i_I= 1$), negative ($\eta^i_I=- 1$) or vanishing charge ($\eta^i_I=0$) under $A^i$. We can constrain the tension and charges from effective field theory considerations and, of course, the WGC.  For the $i$th membrane species to be part of the effective field theory, we require 
\be
q_i \lesssim \muv^2, \qquad \tau_i \lesssim \muv^3. \label{EFT}
\ee
Furthermore, there ought to be at least one species for which we also satisfy the constraints coming from the  WGC \cite{Arkani-Hamed:2006emk,Ibanez:2015fcv}
\bea
\tau_i<M_{pl} q_i \qquad &\text{electric WGC} \label{WGC1} \\
\muv^3< q_i M_{pl}\qquad &\text{magnetic WGC\ . \label{WGC2}}
\eea

To make contact with our previous work \cite{us}, we switch to a dual formulation,  introducing a Lagrange multlplier $\phi_i$ fixing $F^i=dA^i$ on shell, via a term $\int \phi_i (F^i-dA^i)$. Integrating out the scalars, $\star F_i$, yields
\be\label{Lorentzianactionf}
S =\int_\mathcal{M} d^4 x\sqrt{|g|} \left[\frac{M_{pl}^2}{2}  R-L_f(\phi_i) \right]-\int \phi_i dA^i
 +S_\text{boundary}+S_\text{membranes}, 
\ee
where $L_f$ is the Legendre transform of $f$. This action now falls into the general class studied in \cite{us}, with $Z_{ij}=\omega^{ij}=0$, $\sigma_i=-\phi_i$ and $V(\phi)=L_f(\phi)$.  As we are interested in transitions between  eigenstates of constant  four-form flux, we assume Neumann boundary conditions on the three-form fields but Dirichlet boundary conditions for the metric and the dual scalars. Correspondingly, the boundary part of the action is given by \cite{Gibbons:1976ue,Duncan:1989ug, us}
\be
S_\text{boundary}=M_{pl}^2 \int_{\partial \mathcal{M}}d^3 \xi \sqrt{|\gamma|}   K  -\int_{\partial \mathcal{M}}  \chi_i\alpha^i, 
\ee     
where $K=\gamma^{ab}K_{ab}$ is the trace of the extrinsic curvature,  $K_{ab}=\frac12 \mathcal{L}\gamma_{ab}$, defined as the Lie derivative of the induced metric with respect to the outward normal $n^\nu$. The  momenta conjugate to the 3-forms are given by $\chi_i=-\phi_i$.

The vacua are real Lorentzian solutions with constant scalars,  $\phi_i$, four-forms of constant flux $\star F^i=c^i$, and a maximally symmetric  metric with constant curvature $k^2$, corresponding to   de Sitter ($k^2>0$),  Minkowski ($k^2=0$) or anti de Sitter ($k^2<0$) spacetime satisfying
\be
3 M_{pl}^2 k^2 = \vqft+L_f(\phi) , \qquad 
\frac{\del L_f}{\del \phi_i}=c^i, 
\ee
where we have included the contribution from  the underlying vacuum energy, $\vqft$, explicitly. The conjugate momenta $\chi_i=-\phi_i$ are locally constant and, in the presence of membranes, quantised in units of the membrane charge, yielding  
\begin{equation}
\phi_i=N_i q_i  \qquad \text{(no sum)}\ .
\end{equation}
At the membranes,  we must specify junction conditions. The metric and  three-forms  are required to be continuous in order for the membrane action to be well defined. However, their normal derivatives can be discontinuous. Indeed, the standard Israel junction conditions \cite{Israel} relate the jump in the  extrinsic curvature of the brane to its tension. The momentum conjugate to the three-forms also jumps, giving 
\be \label{chijump}
\Delta \phi_i=\eta^i_I q_i \qquad  (\text{no sum over $i$}).
\ee
It follows  that the flux number, $N_i$, changes by a single unit as we move across a membrane charged under $A^i$. 

Transitions between vacua are therefore mediated by membrane nucleation. To compute the rate at which membranes are nucleated,  and transitions occur, we analytically continue to Euclidean signature, and solve for the instanton solution interpolating between the parent vacuum, $\mathcal{M}_+$, with curvature $k_+^2$ and the daughter vacuum, $\mathcal{M}_-$, with curvature $k_-^2$.  We analytically continue as follows
\be
t \to -i t_E, \qquad \star F^i \to \star F^i, \qquad A^i\to iA^i, \qquad S \to iS_E,
\ee
where $S_E$ is the Euclidean action and solve for $O(4)$ symmetric Euclidean field configurations, with metric
\be
ds^2=dr^2+\rho(r)^2 d \Omega_3,
\ee
where $d \Omega_3=h_{ij} d\xi^i d\xi^j$ is the metric on a unit 3-sphere, Euclidean three-form potentials 
\be
A^i=A^i(r) \sqrt{|h|} d^3 \xi,
\ee
and scalars $\phi_i=\phi_i(r)$. The radial coordinate is assumed to run from $r_\text{min}<0$ to $r_\text{max}>0$ and the membrane lies at $r=0$.   Details of the corresponding field equations and their solutions are presented in \cite{us}.  Note that upon Wick rotation back to the Lorentzian signature, the instanton solution corresponds to a bubble of daughter vacuum in the parent spacetime. 

In semi-classical theory of vacuum decay, the transition rates between vacua $\mathcal{M}_+ \to \mathcal{M}_-$ are given by \cite{Coleman:1977py, Callan:1977pt, Coleman:1980aw}
\be
\frac{\Gamma}{\text{Vol}} \sim e^{-B/\hbar},
\ee
where $\Gamma$ is the transition rate and the tunnelling exponent
\be
B=S_E(\text{instanton})-S_E(\text{parent}).
\ee
Here $S_E(\text{instanton})$ is the Euclidean action evaluated on the bubble configurations described above, interpolating between the vacua $\mathcal{M}_+$ and $\mathcal{M}_-$.  In contrast, $S_E(\text{parent})$ is the Euclidean action evaluated on the complete parent vacuum,  $\mathcal{M}_+$, with no bubbles. After a lengthy calculation, we find \cite{us}
\be \label{B}
B=\frac{4M_{pl}^2 \Omega_3}{k_+^2 } \left[\frac{1+Y-X}{Y(1+Y+X)} \right]\, ,
\ee
where we define two important parameters 
 \be
X=\frac{4M_{pl}^4 \Delta k^2}{T^2}, \qquad  Y(X)=\sqrt{(X-1)^2+16 k_+^2 M_{pl}^4/T^2},
 \label{eq:defY}
 \ee
where $\Delta k^2=k_+^2-k_-^2$ is the jump in the vacuum curvature and $T$ is the tension of the brane mediating the transition. 

Transitions between vacua are readily  constrained on physical grounds. In particular, we only allow geometric configurations  supported by non-negative membrane tensions. We also require that the tunnelling exponent is finite so that the transition is not infinitely suppressed.  This allows  for three configurations of physical interest
\begin{itemize}
    \item $\text{dS}_+ \to \text{dS}_-$
    \item $\text{dS}_+ \to \text{Minkowski/AdS}_-$
    \item $\text{Minkowski/AdS}_+ \to \text{Minkowski/AdS}_-$  $(|k_-| \geq |k_+|)$
\end{itemize}
In what follows, we will focus on the case where the parent vacuum is de Sitter and the transition occurs to a vacuum of smaller curvature. Given our presence in a  very low curvature vacuum excited by matter and radiation, we are  particularly interested in two types of transition: rapid decay from high or intermediate curvature to the current low curvature vacuum, required to avoid the empty universe problem;  decay from the current vacuum into an apocalyptic anti de Sitter universe.  A detailed study of the decay dynamics follows in section \ref{sec:dsdecay}.

For now, we focus on the latter scenario and compute the rate of transition from a near Minkowski vacuum to anti de Sitter. In this limit, the corresponding tunnelling exponent goes as \cite{us}
\begin{align} \label{BMAdS}
B_{M_+ \to AdS_-} & \sim  \frac{2 M_{pl}^2 \Omega_3}{k_+^2}(1-S(X))\\ \nonumber
& +\frac{8 M_{pl}^6 \Omega_3}{T^2 X(X-1)^2}\left[(X-1)^2(1-S(X))+2S(X)\right],
\end{align}
where $X_{M_+ \to AdS_-}\approx -\frac{4M_{pl}^4 k_-^2}{T^2}$ and $S(X)=\text{sgn}(X-1)$. Here we see the presence of a possible pole in the bounce as the curvature of the parent vacuum approaches zero, $k_+^2 \to 0$.  The existence of this pole would ensure the stability of a Minkowski vacuum and the exponential longevity of very low scale de Sitter vacua. The pole is present when the parameter $|X_{M_+ \to AdS_-}| <1$ and otherwise absent.  Thus, in order to halt or emphatically slow down the rate of membrane nucleation, as described in the introduction,  we require $|X_{M_+ \to AdS_-}| <1$ \cite{us}.

\section{Implications of WGC on specific models} \label{sec:models}

Let us now explore the impact of the WGC on the set-up described in the previous section. To this end,  it will be convenient to express the spacetime curvature in terms of the dimensionless flux $Q_i=N_i \delta_i$ for $\delta_i= q_i/\muv^2$, via an expression of the form
\be
3 M_{pl}^2 k^2=\vqft+ \muv^4 \kappa(Q), \label{kappadef}
\ee
where $\kappa(Q)=L_f(\muv^2 Q)/\muv^4$ is a dimensionless function of the flux. Note that if the membranes of type $i$ are inside the EFT, we have $\delta_i <1$, as per equation \eqref{EFT}.  Now suppose we nucleate such a membrane, so that $N_i \to N_i \pm 1$ and $N_j \to N_j $ for $j\neq i$.  The process induces a corresponding change in $k^2$,
\be
\Delta k^2=\frac{\muv^4}{3 M_{pl}^2} \Delta \kappa_i
\ee
where we define
\be \label{Dk}
\Delta \kappa_i\equiv \kappa( \ldots, Q_{i-1}, Q_i, Q_{i+1}, \ldots)-\kappa(\ldots, Q_{i-1}, Q_i \pm \delta_i, Q_{i+1}, \ldots)=-\sum_{n=1}^\infty \frac{(\pm \delta_i)^n}{n!} \frac{\del^n }{\del Q_i^n}\kappa(Q)\ ,
\ee
and in the last equality we Taylor expanded $\kappa(\ldots, Q_{i-1}, Q_i \pm \delta_i, Q_{i+1}, \ldots)$.

As we described in section \ref{sec:review}, the nucleation rate is controlled by the value of a parameter
\be
X_i=\frac{4 M_{pl}^4 \Delta k^2}{\tau_\is^2}=\frac{4 M_{pl}^2 \muv^4}{3\tau_i^2}\Delta \kappa_i\ .
\ee
For a parent vacuum with $k_+^2\approx 0$,  we must choose the flux such that $\kappa  \approx  -\frac{\vqft}{\muv^4}$. If we further assume that $|N_\is| \gg 1$ when $k^2\approx 0$,  it follows that $|Q_i| \gg \delta_i$ where we can truncate the Taylor expansion to leading order in \eqref{Dk} to show that 
\be
\left|X_i^{M_+ \to AdS_-} \right| \approx \frac43 \left(\frac{q_i M_{pl}}{\tau_i}\right)^2 \frac{\muv^2}{q_i}  \left|\frac{\del \kappa}{\del Q_i}\right|_{k^2=0} \ .
\ee
For membrane nucleation to be halted at very low curvature, we require    $\left|X_i^{M_+ \to AdS_-} \right|<1$, or equivalently 
\be
\left|\frac{\del \kappa}{\del Q_i}\right|_{k^2=0}<\frac{3}{4}\left(\frac{\tau_i}{q_i M_{pl}}\right)^2 \frac{q_i}{\muv^2}
\label{eq:flatk}
\ee
for {\it all} species. If the membranes of type $i$ are part of the EFT \eqref{EFT} and satisfy the WGC \eqref{WGC1}, it is clear that the corresponding first partial derivative of $\kappa$ should not be too large near the $k^2 \approx 0$ vacuum.  We will now explore the implications of this for several models.

\subsection{Homogeneous polynomial models}

In the Kaloper-Westphal (KW) model \cite{Kaloper:2022jpv},  the cosmological constant is linear in the flux, $\kappa=a^i Q_i$. This represents a multi-field generalisation of Hennueaux and Teitelboim's covariant formulation of unimodular gravity (with branes). It cannot be obtained  from a generic action of the form \eqref{Lorentzianaction}, since the Legendre transform cannot be inverted in the linear case. Even so, it is instructive to explore how the WGC may be avoided even when the halting condition \eqref{eq:flatk} holds for near Minkowski vacua.  In the KW model, the  $q_i$ have irrational ratios, so we can always find a solution to $k^2 \approx 0$ to any desired level of accuracy as long as we choose sufficiently large $|N_i|$.  Descent between vacua is halted at very low curvatures provided
 \be \label{ai}
 |a^i|< \frac{3}{4}\left(\frac{\tau_i}{q_i M_{pl}}\right)^2 \frac{q_i}{\muv^2}
 \ee
 for all values of $i$. Clearly, if there are any species satisfying WGC \eqref{WGC1} along with the EFT constraint \eqref{EFT}, the corresponding value of $a^i$ should be tuned to be suitably small.  
 
When the cosmological constant is quadratic in flux as in the Bousso-Polchinski model difficulties can arise, as was first noted in \cite{us} and emphasized in \cite{Kaloper:2023vrs}. To see this note that $\kappa(Q)=\frac12 \sum_{i} Q_i^2$ where the coefficients are fixed by  canonical normalisation $a_{ij}=\delta_{ij}$. The halting  condition \eqref{eq:flatk} now requires
 \be
 |Q_i| <\frac{3}{4}\left(\frac{\tau_i}{q_i M_{pl}}\right)^2 \frac{q_i}{\muv^2} \implies |N_i|  <\frac{3}{4}\left(\frac{\tau_i}{q_i M_{pl}}\right)^2
 \ee
for all values of $i$.  The flux numbers are  expected to be large in a neighbourhood of the Minkowski vacuum in order to achieve the necessary cancellation of the presumably large bare vacuum energy. This is clearly incompatible with any species satisfying the WGC \eqref{WGC1}. 

What is happening here? The point is that in order to halt the descent between vacua at very low curvatures the flux dependent function $\kappa(Q)$ is required to be flat along any direction satisfying the WGC, as per the inequality \eqref{eq:flatk}. In the case of the Bousso Polchinski model, the function $\kappa(Q)$ is just a sum of squares, being flatter near their minimum at $Q_i=0$. However, in the generic case where $\vqft \gg M_{pl}^2 H_0^2$, this is not a low curvature vacuum. 

We can  generalise this result to the case where $\kappa(Q)$ is a homogeneous function of any degree $n  > 1$.  Indeed, in this case, the Euler identity implies that $\sum_i Q_i \frac{\del \kappa}{\del Q_i}=n 
\kappa$, from which we infer
\be
\frac{\del \kappa}{\del Q_i}=\frac{1}{n-1} \sum_j Q_j \frac{\del^2 \kappa}{\del Q_i \del Q_j}.
\ee
Plugging this result back into the inequality  \eqref{eq:flatk}, we see that the halting condition requires
\be
\left| \sum_j N_j \frac{q_j}{q_i}  \frac{\del^2 \kappa}{\del Q_i \del Q_j} \right|_{k^2=0}  <\frac{3}{4}\left(\frac{\tau_i}{q_i M_{pl}}\right)^2 (n-1) \label{flatHess}
\ee
for all values of $i$. Recall that the $|N_i|$ are generically expected to be large on the Minkowski vacuum.  Assuming the absence of any charge hierarchies $q_i \sim \mathcal{O}(q)$, there are two ways in which the inequality \eqref{flatHess} might be satisfied alongside the WGC  \eqref{WGC1}: either there are miraculous cancellations between terms or   certain components of the Hessian turn out to be extremely small.  Note that the inequality \eqref{flatHess}  must hold even if there are order one corrections to the underlying vacuum energy. In a generic scenario, this is likely to spoil any miraculous cancellations, so we are led  to assume fine tuning of the Hessian. Indeed, if we suppose that the first species with $i=1$ satisfies the WGC, we must have that 
\be
\left| \frac{\del^2 \kappa}{\del Q_1 \del Q_j} \right|_{k^2=0} \ll \frac34 (n-1)
\ee
for all $j$. If more species satisfy the WGC, then more components of the Hessian will need to be tuned. 

For a quadratic model with $n=2$, this constraint on the Hessian requires the corresponding couplings to be unnaturally small. This is consistent with our previous result for the Bousso-Polchinski model where we saw that the halting condition for low curvature vacua  was incompatible with the weak gravity conjecture \cite{us}.  However, for higher powers, $n>2$, the Hessian can be rendered small by taking the $Q_i$ to be small in the corresponding limit. For a suitable choice of membranes, this may still be possible even for large $N_i$.  

To see this explicitly, let us consider the case where $\kappa$ is dominated by a sum of monomials of degree $n$, 
\be
\kappa(Q) = \sum_i \frac{c_n^i}{n!} Q_i^n 
\ee
where $n \geq 1 $ and $c_n^i$ are constants, assumed to be order one in a natural scenario.  Recall that the nucleation of the membrane of type $i$ triggers a change in flux  $N_i\rightarrow N_i\pm 1$ and $N_{j\neq i}\rightarrow N_{j\neq i}$. The corresponding value of $X_i$ controlling the rate of the transition is given by 
\be
|X_i| \approx \frac{4}{3}n \left(\frac{q_i M_{pl}}{\tau_i}\right)^2 \frac{\muv^2}{q_i}  \frac{|c_n^i|}{n!} |Q_i|^{n-1}
\ee
where we have assumed that the flux number is large $|N_i| > 1$. In the absence of any hierarchies, $q_i \sim \order(q)$, $\tau_i \sim \order (\tau)$ and $c_n^i \sim \order (c_n)$. Near the Minkowski vacuum, we have $\kappa  \approx  -\frac{\vqft}{\muv^4}$, and since we expect each $|Q_i| \sim   |n! \kappa/c_n|^\frac1n$
 we infer that 
\be \label{Ni}
|N_i| \sim \frac{\muv^2}{q} \left| \frac{\vqft n!}{\muv^4 c_n}\right|^{\frac1n}
\ee
and 
\be \label{X}
\left|X_i^{M_+ \to AdS_- }\right| \sim   \frac43  \left(\frac{q M_{pl}}{\tau }\right)^2 \frac{\muv^2}{q}   \left|\frac{\vqft}{\muv^4}\right|^{1-\frac1n} \left(\frac{n^n}{n!}\right)^{\frac1n} |c_n|^\frac1n \ .
\ee
Recall that generically we expect $|N_i| >1$. For the descent between vacua to halt at very low curvatures,   we also require $\left|X_i^{M_+ \to AdS_- }\right|<1$ for each species.  From equation \eqref{X}, we  see that this is only possible if there are one or more of the following:
\begin{itemize}
    \item a violation of the WGC, $\frac{q M_{pl}}{\tau}<1 $
    \item the membrane charges are no longer consistent with the EFT,  $\frac{\muv^2}{q} < 1$
    \item suppression of the couplings, $|c_n|<1$
    \item suppression of the bare cosmological constant: $\frac{|V_{QFT}|}{M_{UV}^4}<1$
\end{itemize}
None of these scenarios are ideal but perhaps the least worrying is the suppression of the bare cosmological constant.  Indeed, if the scale of supersymmetry breaking lies many orders of magnitude below the cut-off, this is exactly what we would expect to find. 

Note that we can recast the large flux condition $|N_i|>1$ and the halting condition,  $\left|X_i^{M_+ \to AdS_-} \right|<1$, as the following constraint on the couplings
\be \label{cineq}
|c_n| < \text{min}\{ \lambda_1, \lambda_2\}
\ee
where 
\be
\lambda_1=\left(\frac{\muv^2}{q}\right)^n   \frac{|\vqft|}{\muv^4 } n!, \qquad \lambda_2 =\left( \frac{3}{4} \left(\frac{\tau}{q M_{pl}}\right)^2 \frac{q}{\muv^2}   \right)^n \left(\frac{|\vqft|}{\muv^4 } \right)^{1-n} \frac{n!}{n^n}
\ee
If we now demand that both $\lambda_1>1$ and $\lambda_2>1$, as required to avoid fine tuning of the couplings, we find that
\be
\frac{3}{4} \left(\frac{\tau}{q M_{pl}} \right)^2> \frac{1}{(n-1)!} \left( \frac{q}{\muv^2}\right)^{n-2}
\ee
For $n=2$, it is clear that this is not consistent with the weak gravity conjecture, as we have already seen. However, for $n>2$, we can be more optimistic. Indeed, as an example, consider the case where $n=10$,  $|\vqft| \sim 10^{-6} M_{pl}^4 $, $q \sim 10^{-2} M_{pl}^2$, $\tau \sim  10^{-3} M_{pl}^3$, $|c| \sim 1$ and $\muv \sim M_{pl}$, so that
\be
|N_i| \sim 114, \qquad \left|X_i^{M_+ \to AdS_-} \right| \sim 0.1 
\ee
In this scenario, the membrane charges and tensions are chosen to be small but are consistent with the EFT constraint \eqref{EFT} and the WGC \eqref{WGC1}. Similarly, the bare cosmological constant is many orders of magnitude below the cut-off, as one might expect in a supersymmetric scenario.  Nevertheless, for order one couplings in $\kappa(Q)$, it seems we can have scenarios in which membrane nucleation is halted at very low curvatures, without any violation of our EFT constraints or the WGC. 

Of course, in a consistent weakly coupled scenario we would not expect higher order operators to dominate over the quadratic terms. Given that we have already seen that canonical quadratic terms are incompatible with the halting condition if the WGC  and EFT constraints are satisfied, it follows that we can only exploit the higher order operators by going to strong coupling.  Further, in such a scenario, we would expect a whole tower of higher order operators to be present  beyond the homogeneous polynomials considered  here.  What effect do they have on the interplay between the WGC and the longevity of low curvature vacua? We shall now explore this question in two different ways: first, with a generic tower of EFT operators whose coefficients are set using naive dimensional analysis (NDA) \cite{Manohar:1983md,Gavela:2016bzc}; and second, in the special case of a generalised DBI action for four-forms, where the couplings are motivated by the dynamics of a spacetime filling 3-brane \cite{Jurco:2012yv,Ho:2014una}.

\subsection{A generic tower of higher order operators} 
Consider a generic tower of EFT operators whose coefficients are set using naive dimensional analysis \cite{Manohar:1983md,Gavela:2016bzc}. This allows us to be a little more precise about the form of the dimensionless coefficients in the dual potential in equation \eqref{Lorentzianactionf}.  In particular, we expect that \cite{DAmico:2017cda,Padilla:2018hvp}
\be
L_f(\phi)=\sum_{n \geq 2}  \frac{b^{i_1 \ldots i_n}}{ (\muv^2/4\pi)^{n-2} n!} \phi_{i_1} \ldots \phi_{i_n}  
\ee
where $b^{ij}=\delta^{ij}$ and the $b^{i_1 \ldots i_n} \sim \order(1)$.  In terms of the dimensionless potential, this gives
\be
\kappa(Q)=\sum_{n \geq 2}  \frac{b^{i_1 \ldots i_n} }{ n!}  (4\pi)^{n-2} Q_{i_1} \ldots Q_{i_n}  
\ee
In the weakly coupled regime, $|Q_i|< \frac{1}{4 \pi}$, the leading order quadratic terms dominate and we recover the negative results of the previous section. However, the EFT is valid even in the strongly coupled regime where $\frac{1}{4 \pi}<|Q_i|<1$, allowing us to go beyond the leading order quadratic approximation.  To see what happens then, consider the case where  $|Q_i| \sim \order(Q)$ in the near Minkowski limit, where the scale $Q$ is assumed to lie in the interval,  $\frac{1}{4 \pi}<|Q|<1$.  Recall that the lower limit means we are at strong coupling and cannot truncate the expansion in the $Q_i$, while the upper limit ensures that the EFT remains valid.  We  then find that\footnote{For example, this result is easily understood for the  case where the dimensionless potential scales as $\kappa \sim e^{4 \pi Q}/(4\pi^2)$ in the strong coupling regime. }
\be\left|\frac{\partial\kappa}{\partial Q_i}\right|_{k^2=0} \sim 4 \pi |\kappa|_{k^2=0} \sim 4 \pi \frac{|\vqft |}{\muv^4}
\ee
Near the Minkowski vacuum  $|\kappa|_{k^2=0} \approx \frac{|\vqft |}{\muv^4}$  and so we expect $ Q \sim \frac{1}{4 \pi}\ln(4 \pi |\vqft|/\muv^4)$. We then infer that the flux numbers scale as 
\be
|N_i| \sim \frac{\muv^2}{q} Q \sim \frac{\muv^2}{4 \pi q} \ln(4 \pi |\vqft|/\muv^4)
\ee
while
\be
\left|X_i^{M_+ \to AdS_-} \right| \sim \frac{16 \pi}{3} \left(\frac{q M_{pl}}{\tau}\right)^2 \left(\frac{\muv^2}{q}  \right)\frac{|\vqft |}{\muv^4}
\ee
where we have, once again, assumed the absence of any hierarchies between species, so that $q_i \sim \order(q)$ and $\tau_i \sim \order(\tau)$.  If the WGC is to be respected  and the membranes stay inside the EFT, we might hope to satisfy the halting condition, $\left|X_i^{M_+ \to AdS_-} \right| <1$,  by taking the bare vacuum energy to be well below the cut-off, as we did in the previous section.  The problem is that we expect the scale $Q$ to be in the strong coupling window, $\frac{1}{4 \pi}<|Q|<1$, constraining $|\vqft|\gtrsim \muv^4 e/4\pi$. It follows that 
$$\left|X_i^{M_+ \to AdS_-} \right| \gtrsim  \frac{e}{3} \left(\frac{q M_{pl}}{\tau}\right)^2 \left(\frac{\muv^2}{q}  \right) \sim \frac{e}{3} \left(\frac{q M_{pl}}{\tau}\right)^2 \frac{|N_i|}{Q}>\frac{e}{3} \left(\frac{q M_{pl}}{\tau}\right)^2 |N_i|  $$ 
Given that flux numbers are always assumed to be large, there is no way this can be less than unity without violating the WGC. This suggests that generic strong coupling scenarios do not help alleviate tensions between the WGC and our ability to halt the descent between vacua at very low curvatures. Of course, there may be some theories where the desired flattening of  $\kappa(Q)$ is achieved at strong coupling thanks to elegant cancellations between terms, perhaps motivated by some enhanced symmetry. We will consider exactly such a scenario in the next section.

\subsection{Generalised DBI}
We now consider the generalised DBI action for three-form fields \cite{Jurco:2012yv} (see also \cite{Ho:2014una} for a more transparent description of the action) with multiple species, so that the action is given by equation \eqref{Lorentzianaction}  with
 \be
 f(\star F_i)=\sum_i \Lambda_i^4 \left(1-\sqrt{1-\frac{(\star F^i)^2}{\Lambda_i^4}}\right)
 \ee
This can be seen as the action for a spacetime filling 3-brane with tension $\Lambda_i^4$ on whose world volume the $F_i$ are defined.
 The Legendre transform is easily computed to give
 \be
 L_f(\phi)=\sum_i \Lambda_i^4 \left( \sqrt{1+\frac{\phi_i^2}{\Lambda_i^4}}-1\right)
 \ee
 with a corresponding dimensionless potential
 \be
 \kappa (Q)=\sum_i \lambda_i^4 \left( \sqrt{1+\frac{Q_i^2}{\lambda_i^4}}-1\right), \qquad \lambda_i=\frac{\Lambda_i}{\muv}
 \ee
 To avoid falling into the trap of quadratic domination and problems with the WGC, let us assume that when $k^2 \approx 0$,  we have $|Q_i| \gtrsim \lambda_i^2$.  In this limit, we recover the linear behaviour familiar from the KW model, albeit from a more well motivated initial set-up
 \be
 \kappa (Q) \approx \sum_i \lambda_i^2 |Q_i |\ .
 \ee
Since this is always positive, it is clear that we must have $\vqft<0$ in order to have  a  spectrum of vacua that includes very low curvatures. For near Minkowski vacua, let us further assume that $|Q_i| \sim \order(Q)$ for some scale $Q$ and $\lambda_i \sim \order(\lambda)$ for some scale $\lambda$, so that  $Q \sim  |\vqft |/\muv^4 \lambda^2 \mathcal{N} $ where $\mathcal{N}$ is the number of species.  It also follows that $|\partial\kappa/\partial Q_i|_{k^2=0}  \approx  \lambda_i^2$. For the EFT constraints \eqref{EFT} and the WGC \eqref{WGC1} to be satisfied alongside the halting condition \eqref{eq:flatk}, we know that we must have a flat potential, and so $\lambda_i^2<1$, or equivalently, the underlying brane tensions lying below the cut-off $\Lambda_i < \muv$. This is certainly what we would expect from a consistent EFT. 

We now consider the density of vacua  at low curvature for the generalised DBI model. If we have just two species of three-form field, we can adopt the KW strategy and assume an irrational charge ratio.  This guarantees a dense set of vacua  even  when the membrane charges are not especially small, allowing us to avoid the empty universe problem.  Alternatively, we can consider the generalised DBI framework as a UV completion of the Bousso-Polchinski model, with a large number of fields.  To read off the density of vacua at large $\mathcal{N}$ and large $Q_i=N_i q_i/ \muv^2 \gg  \lambda_i^2$,  we note that the curvature 
\be
k^2 \approx \frac{\vqft+ \sum_i \Lambda_i^2 q_i |N_i|}{3 M_{pl}^2}.
\ee
This means the vacua span an $\mathcal{N}$ dimensional grid with spacing $\Lambda_i^2 q_i$. Surfaces of constant vacuum curvature correspond to the boundaries of  cross-polytopes, as opposed to spheres in the case of the original Bousso-Polchinski model. 
Consider two such surfaces of curvature $k^2=\frac{\vqft+r}{3 M_{pl}^2}$ and  $k'^2=\frac{\vqft+r'}{3 M_{pl}^2}$, where $r'>r>0$. The corresponding polytopes differ by a shell of volume $V_\text{shell}=\frac{2^{\mathcal{N}}} {\mathcal{N}!}(r'{}^\mathcal{N}-r^\mathcal{N})\approx \frac{ 2^\mathcal{N} r^{\mathcal{N}-1}} {(\mathcal{N}-1)!}(r'-r)$. To ensure that this shell contains at least one grid point, we require that $V_\text{shell} \gtrsim D \prod_{i=1}^\mathcal{N} \Lambda_i^2 q_i$, where $D$ is the degeneracy factor. This implies that vacua are separated by a curvature
\be
\delta k^2 \approx \frac{ (\mathcal{N}-1)!  D \prod_{i=1}^\mathcal{N} \Lambda_i^2 q_i}{3 M_{pl}^2  2^\mathcal{N} r^{\mathcal{N}-1} }, \qquad r=3M_{pl}^2 k^2-\vqft
\ee
Assuming the $\Lambda_i^2 q_i$ are incommensurate, the degeneracy $D=2^\mathcal{N}$ due to the $N_i \to -N_i$ symmetry. The density of vacua near Minkowski space is therefore given by
 \be
\delta k^2|_{k^2=0} \approx \frac{ (\mathcal{N}-1)!   \prod_{i=1}^\mathcal{N} \Lambda_i^2 q_i}{3 M_{pl}^2  |\vqft|^{\mathcal{N}-1} },
\ee
In the absence of hierarchies in membrane charge and DBI tension, we assume that $q_i \sim \order(q)$ and $\Lambda_i \sim \order(\Lambda)$, so that 
 \be
\delta k^2|_{k^2=0} \sim \frac{ (\mathcal{N}-1)!  ( \Lambda^2 q)^\mathcal{N}}{3 M_{pl}^2  |\vqft|^{\mathcal{N}-1} } \sim  \frac{ |\vqft| }{3 M_{pl}^2 \mathcal{N} }  \left(\frac{\Lambda^2 q  }{|\vqft| }\right)^\mathcal{N}\mathcal{N}! 
\ee
Using Stirling's approximation $\mathcal{N}!  \approx \sqrt{2 \pi \mathcal{N} } \left(\frac{\mathcal{N}}{e} \right)^\mathcal{N}$ for large $\mathcal{N}$ we have that 
$$\delta k^2|_{k^2=0} \sim \frac{ |\vqft| }{3 M_{pl}^2 } \sqrt{\frac{2\pi}{\mathcal{N}}} \left(\frac{\mathcal{N}\Lambda^2 q}{|\vqft| e} \right)^\mathcal{N}.$$
Requiring $\delta k^2 |_{k^2=0} \lesssim H_0^2 $ only imposes a constraint 
\be
\sqrt{\frac{2\pi}{\mathcal{N}}} \left(\frac{\mathcal{N}\Lambda^2 q}{|\vqft| e} \right)^\mathcal{N} \lesssim \frac{3 M_{pl}^2 H_0^2}{|\vqft|}\ .
\ee
From our analysis of the linear model and equation \eqref{ai},  descent between vacua is halted at low curvature provided
\be
\lambda_i^2 < \frac{3}{4}\left(\frac{\tau_i}{q_i M_{pl}}\right)^2 \frac{q_i}{\muv^2}.
\ee
In the absence of hierarchies this translates into the following constraint on the tensions $\tau_i \sim \order (\tau)$,  
\be
\tau \gtrsim 2M_{pl} \Lambda \sqrt{ \frac{  q}{3}}\ .
\ee
Finally, recall that the EFT constraints \eqref{EFT}, the electric WGC \eqref{WGC1} and the magnetic WGC \eqref{WGC2}  require that
\be
\tau \lesssim M_{pl} q, \qquad \frac{\muv}{M_{pl}} \lesssim \frac{q}{\muv^2} \lesssim 1 \ .
\ee
All constraints can be satisfied if we take, for example, 
$|\vqft| \sim M_{pl}^4$, $\mathcal{N} \approx 108$, $\tau \sim 0.1 M_{pl}^3$, $q \sim 0.2 M_{pl}^2$, $\Lambda \sim 0.1 M_{pl}$ and $\muv \sim 0.5  M_{pl}$. Note that this implies that $Q \sim  |\vqft |/\muv^4 \lambda^2 \mathcal{N} \sim 3.7  \gg \lambda^2 \sim 0.04$, which confirms that we are indeed in the asymptotic regime of the DBI action. 

This  suggests that the generalised DBI set-up provides an elegant completion of the Bousso-Polchinski model, capable of halting the descent between vacua at low curvatures without running into any problems with the WGC or the consistency of the EFT.  The model recovers the dynamics of a linear model in the asymptotic regime, with the required suppression of the effective coupling understood in terms of the underlying brane tensions lying inside the cut-off.

\section{de Sitter decay dynamics} \label{sec:dsdecay}
As we have seen, membrane nucleation allows for descent from high scale de Sitter vacua to the low scale vacuum we see today, and perhaps even beyond.  The dynamics of this descent is controlled by the tunnelling exponent given in equation \eqref{B}. Our goal here is to investigate how this changes as we move through the landscape, with a view to better understanding how the landscape is populated and how we arrive at the current vacuum.

The first thing to note is that for a parent de Sitter or Minkowski vacuum, $k_+^2 \geq 0$, the tunnelling exponent is always positive and monotonically decreasing with $X$, as already emphasized in \cite{us}.  
To delve a little deeper, we note that the behaviour of the tunnelling exponent depends both on the curvature of the parent vacuum through the dimensionless combination $k_+^2 M_{pl}^4/T^2$ and on the magnitude of the curvature jumps  between successive vacua through $X=4M_{pl}^4 \Delta k^2 /T^2$. Therefore curvature and its jumps between vacua are classified as large or small with respect to the reference  scale $T^2/M_{pl}^4$. This observation prompts us to analyse equation \eqref{B} in the asymptotic limits $k_+^2\gg T^2/M_{pl}^4$  and $k_+^2\ll T^2/M_{pl}^4$.  We focus on the case of vacuum descent, so that $\Delta k^2>0$ and by association, $X>0$. 

\begin{figure}
    \centering
\includegraphics[width=0.7\textwidth]{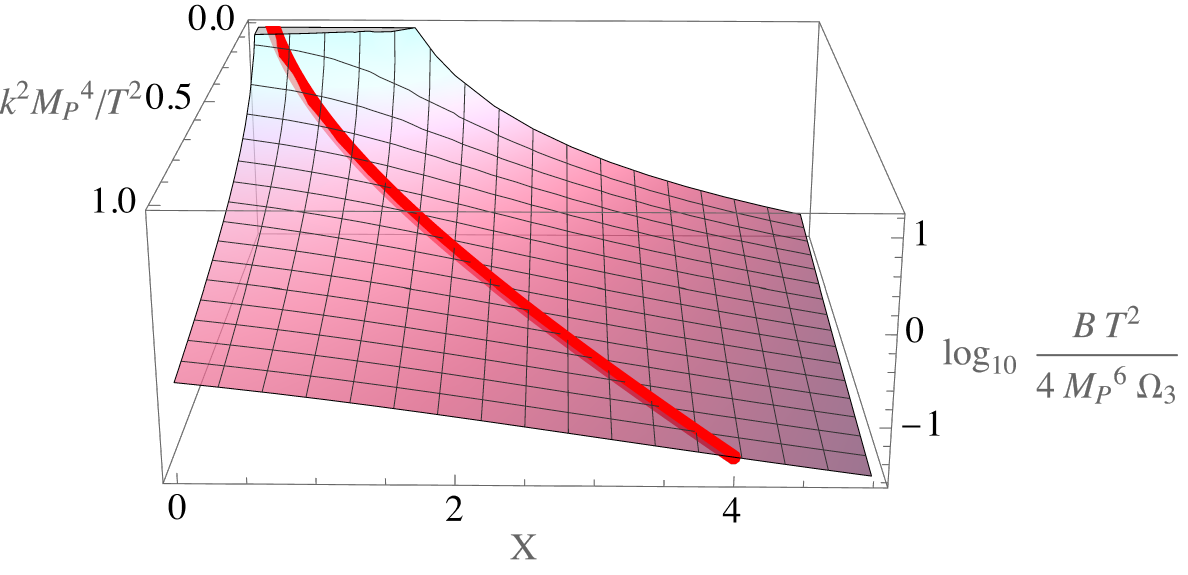}
    \caption{The logarithm of the tunnelling exponent, $\log_{10} (B T^2/4 M_{pl}^6 \Omega_3)$ as a function of  $k_+^2 M_{pl}^4/T^2$ and $X$. Note that as the curvature of the parent vacuum decreases, the value of $B$ increases. }
    \label{fig:B}
\end{figure}

For $k_+^2\gg T^2/M_{pl}^4$ we have that for fixed $X>0$
\be
B\approx \frac{4 M_{pl}^6 \Omega_3}{T^2}\left[ \frac{1}{4}\left(\frac{T^2}{k^2_+ M_{pl}^4}\right)^{3/2}-\frac{X}{8}\left(\frac{T^2}{k_+^2 M_{pl}^4}\right)^2+\mathcal{O}\left(\frac{T^2}{k_+^2 M_{pl}^4}\right)^{5/2}\right]\ .
\label{eq:Bk2large}
\ee
 This approximation is valid in the early stages of vacuum descent when we are in a high scale de Sitter vacuum. At later stages, when the vacuum curvature has fallen to smaller values there are two different scenarios depending on the size of the jump. In  particular, when  $k_+^2\ll T^2/M_{pl}^4$, we find that 
\be \label{eq:BX<1}
B\approx \begin{cases} \frac{4 M_{pl}^6 \Omega_3}{T^2}\left[ \frac{4}{X(X-1)^2}+16 \frac{1-4 X}{X^2 (X-1)^4}\frac{k_+^2 M_{pl}^4}{T^2}+\mathcal{O}\left(\frac{k_+^2 M_{pl}^4}{T^2}\right)^2
\right] & \text{for $X>1$}
 \\
 \frac{4 M_{pl}^6 \Omega_3}{T^2}\left[  \frac{T^2}{k^2_+ M_{pl}^4}+4 \frac{X-2}{(X-1)^2}+16\frac{6-4X+X^2}{(X-1)^4}\frac{k_+^2 M_{pl}^4}{T^2}+\mathcal{O}\left(\frac{k_+^2 M_{pl}^4}{T^2}\right)^2 \right] &\text{for $0<X<1$.}
\end{cases}
\ee
de Sitter decay via flux discharge can now be understood as a model dependent trajectory in the $( k_+^2, \Delta k^2)$ plane, starting at some presumably large curvature $k_+^2$ and moving in the direction in which it decreases.  The changing value of the tunnelling exponent is shown in  figure \ref{fig:B}  as a function of $k_+^2 M_{pl}^4  /T^2$  and $X=4M_{pl}^4 \Delta k^2 /T^2$.
In the limit where $k_+^2 M_{pl}^4  /T^2 \to \infty$ we see from equation \eqref{eq:Bk2large} that $B \to 0$ indicating that very high scale vacua are short lived. This rapid decay of high scale vacua with $B < 1$ was recently dubbed the ``boiling phase"  of vacuum descent \cite{Kaloper:2023vrs}. As the figure shows, as the curvature of the parent vacuum falls, the tunnelling exponent increases and the discharge of the cosmological constant begins to slow down. Eventually we enter the ``braking phase" of vacuum descent  with $B>1$, where the discharge rate becomes exponentially slow \cite{Kaloper:2023vrs}. 

Once the curvature drops below the tension of the membranes, $k_+^2\ll T^2/M_{pl}^6$, we see from equation \eqref{eq:BX<1} that the decay dynamics branches out depending on the size of the jumps in curvature $\Delta k^2$. Indeed, as the curvature of the parent vacuum tends towards zero,  we find that there are two possible asymptotic limits, with the rate of transition from a near Minkowski  vacuum ($k_+^2 \to 0$) to anti de Sitter ($k_-^2<0$) controlled by the following tunnelling exponent,
\be
B_{M_+ \to AdS_-}\sim \begin{cases} \frac{16 M_{pl}^6 \Omega_3}{T^2 X(X-1)^2} & \text{for $X>1$}
 \\
 \frac{4 M_{pl}^6 \Omega_3}{k^2_+ M_{pl}^4} &\text{for $0<X<1$ ,}
\end{cases}
\ee
where now $X=-4M_{pl}^4k_-^2/T^2$. For $X>1$, the tunnelling exponent reaches a finite upper limit, and the corresponding rate of transition settles down to  a finite value depending on the precise value of $X$. Note that the closer $X$ is to unity from above, the slower the rate of transition. We might even say that there is a pole at $X=1^+$ that could be used to  halt any further discharge of the cosmological constant in the Minkowski limit. However, the precise value of $X$ depends on the depth of the would-be daughter vacuum  and the brane tension. In any given model, these quantities are exposed to radiative corrections and any attempt to tune $X$ to be close to unity from above in the appropriate limit would not be natural.

In contrast, for $0<X<1$ we see the presence of the Minkowski pole in the tunnelling exponent, as $k_+^2 \to 0$, indicating that the rate of transition tends towards zero for a range of values of $X$. This is the halting condition that prevents any further discharge of the cosmological constant whenever the vacuum curvature vanishes. By association, low scale de Sitter vacua are extremely long lived.  It follows that in order to achieve a reliable halting mechanism as the parent vacuum approaches Minkowski, we need a model that gives $0<X<1$ in the appropriate limit.

Having established that the discharge of the cosmological constant slows down exponentially at very low curvature provided $0<X<1$, we now ask how we arrive at such a vacuum. In particular, is it possible to ``boil"  directly from a high scale de Sitter vacuum to an exponentially long lived near Minkowski vacuum? Or do we always have pass through an intermediate scale vacuum, with the final transition in the so-called braking regime?

To answer this, consider a transition from a high scale parent vacuum with curvature $k_+^2>0$ to a Minkowski daughter vacuum, with $k_-^2=0$. This has $X_{dS_+ \to M_-}=4M_{pl}^4 k_+^2/T^2$, corresponding to the red line in figure \ref{fig:B}, where the tunnelling exponent is given by
\be 
B_{dS_+ \to M_-}=\frac{16 M_{pl}^6 \Omega_3}{ T^2 X(X+1)^2}\ .
\ee
Let us suppose that this transition is mediated by a membrane of type $i$, with charge $\pm q_i$ and tension, $\tau_i$. Following a similar reasoning to the discussion at the beginning  of section \ref{sec:models}, it follows that the value of $X$ is given approximately by
\be
X^{dS_+ \to M_-}_i \approx \frac{4 M_{pl}^2 \muv^4}{3 \tau_i^2}\left( \frac{\pm q_i}{\muv^2} \right)   \frac{\del \kappa}{\del Q_i}\Big|_{k^2=0} 
\ee
where we recall that $\muv$ is the cut-off and $\kappa(Q)$ is the dimensionless function of the flux defined  through equation \eqref{kappadef}.  As it happens, the halting condition for Minkowski vacua imposes a flatness condition for $\kappa$ along all directions in a neighbourhood of $k^2=0$, as per equation \eqref{eq:flatk}.  Assuming the flux function is differentiable in this  neighbourhood of the Minkowski vacuum, we can use this to bound the value of $X$ describing our transition from de Sitter to Minkowski, giving
\be
\left| X^{dS_+ \to M_-}_i \right| \approx \left| X^{M_+ \to AdS_-}_i \right|<1\ .
\ee
This must be true for all species. In other words, tunnelling from a high scale de Sitter vacuum to a very long lived near Minkowski vacuum must have $0<X_{dS_+ \to M_-} <1$  and so 
\be
B_{dS_+ \to M_-}>\frac{4 M_{pl}^6 \Omega_3}{T^2}\ .
\ee
Since the membranes are assumed to be part of the EFT, we have from \eqref{EFT} that $T<\muv^3 \lesssim M_{pl}^3$. It immediately follows that $B_{dS_+ \to M_-}>1$, or in other words, transitions to exponentially long lived vacua of very low curvature can only occur in the braking phase.  If we assume that we have found ourselves in the current vacuum with any further descent almost halted, we must conclude that we  did not get here by boiling.  

We now arrive at the following picture of our journey through the landscape: the universe starts out in some high scale de Sitter vacuum close to the cut-off and rapidly boils down to some intermediate scale vacua, with curvature still far in excess of the current vacuum. Further descent is much slower. Nevertheless, provided we can ensure $0<X<1$, once a vacuum of very low curvature is nucleated, it lives for a very very long time, far longer than the vacua at intermediate scales.   

As a final comment, we note that in most models there are multiple decay channels corresponding to the nucleation of different membrane species with different values of $X$. However, for fixed values of the curvature in the parent vacuum,   the tunnelling exponent can be shown to be a monotonically decreasing function of $X$ in $X>0$, since
\be
\frac{\del B}{\del X}\Big|_{\text{$k_+^2 $ fixed}}= - \frac{4 M_{pl}^2 k_+^4 B^2 (X+1+2Y)}{\Omega_3 T^2 Y(X-1+Y)^2}
\ee
where we have also used the fact that $Y>0$. This monotonicity implies that largest possible values of $X$ will tend to dominate the transitions. Indeed, consider a low curvature vacuum which can, in principle, decay via instantons with both $X<1$ and with $X>1$.  The latter will dominate since $
B(X<1)>B(X>1)$ for all $k_+^2>0$, leading to a landscape with no preference for small values of the cosmological constant.  Therefore, in order to exploit the halting condition, preventing the decay of Minkowski vacua, it must be that all decay channels satisfy $X<1$, with possible consequences for the WGC across {\it all} species for some models.

\section{Discussion} \label{sec:discuss}
In this paper we have performed a deep dive into the dynamics of models with three-form fields exhibiting discharge of the cosmological constant via membrane nucleation. For suitable choices of parameters, these models can often admit a halting mechanism that stops any further discharge whenever we reach a Minkowski vacuum, presenting a possible solution to the cosmological constant problem. By continuity of the tunnelling rates, it also follows that very low curvature vacua are exponentially long lived in the same parametric regime. As was first noticed in \cite{us}, these parametric choices tend to run into tension with the membrane WGC. 

In \cite{Kaloper:2023vrs} it was argued that the WGC leads to the absence of stable de Sitter space. In this work we have shown that the instability of de Sitter actually arises regardless of the WGC. The existence of charged membranes in the EFT ($\tau<M_\text{UV}^3$) is sufficient to trigger the decay of de Sitter, independently of their charge-to-tension ratio. 

In contrast, the status of Minkowski space is a little more subtle. As we have seen, the  instability of Minkowski space can, in some instances, be connected to membranes satisfying the WGC. For example, in a canonical quadratic model, like the one proposed by Bousso and Polchinski \cite{Bousso:2000xa}, a Minkowski vacuum could only be stable if all species of membrane violate the WGC. Said another way, if just one species of membrane should satisfy the bound set by the  membrane WGC, a Minkowski vacuum would eventually decay. This reminds us of the inevitable decay of charged extremal black holes whenever there is at least one particle whose charge exceeds its mass in Planck units \cite{Arkani-Hamed:2006emk}.

Whilst it is tempting to use the WGC to draw analogies between the stability of Minkowski and/or de Sitter and the stability of extremal black holes, we urge caution in doing so. Indeed, following \cite{Chernyavsky:2017xwm},  we can identify the cosmological constant with conserved charges coming from the global part of the three-form gauge symmetries. More precisely, the conserved charges correspond to the four-form fluxes, $N_i q_i$. In the usual form of the WGC, decay of black hole charge is required to avoid the existence of stable charged remnants. In analogy, we might say that a spacetime can always decay away its charge, so that the only stable solution is the one with vanishing charge. In the Bousso-Polchinski set-up, this corresponds to the anti de Sitter vacuum with vanishing flux and curvature $k^2=\vqft/3 M_{pl}^2<0$. 

We have also shown that tension between the WGC and stability of Minkowski space is  not inevitable and can actually  be avoided in some special cases. In particular, we presented an elegant extension of the Bousso-Polchinski model of multiple three-form species, given by a four-form generalisation of the DBI action.  At small values of the four-form flux, the leading order quadratic terms dominate, yielding the dynamics of the original Bousso-Polchinski model. However, at larger values of the flux, relevant for transitions to and from low curvature vacua, the DBI structure approximates a linear behaviour, albeit in absolute value. In this limit, the theory admits a halting mechanism that picks out the vacuum with the lowest absolute curvature from the landscape, even when we choose natural values for the couplings without any violation of the WGC. Aside from being well motivated, the generalised DBI model also has the property that the effective  potential of the flux is bounded from below, even though it approximates a linear regime at large values of the flux. This is not the case for the original linear flux model \cite{Kaloper:2022jpv} where the potential can become more and more negative for as long as the EFT remains valid. 

Beyond these considerations of the WGC, we have also taken a closer look at how we journey through the landscape of four-form flux vacua, in a relatively model independent way. In particular, our analysis of the dynamics of de Sitter decay demonstrated that the decay rate slows down as the curvature of the parent vacuum decreases.  Decay from high scale de Sitter vacua is rapid, said to be in the so-called boiling phase, whereas decay from low scale de Sitter vacua is much slower, said to be in the so-called braking phase \cite{Kaloper:2023vrs}. The transition from boiling to braking occurs when the tunnelling exponent passes from  $B < 1$ to $B > 1$. With this perspective, how did we arrive at the current low scale vacuum?  The answer depends on whether or not our vacuum is exponentially long lived. If it is, and the halting conditions hold, $|X|<1$, we find that the descent from high scale de Sitter to the current vacuum will not occur through boiling alone - at least one transition must occur in the braking phase.  

Whilst the idea of boiling and braking is a useful one to help visualise the vacuum descent, it doesn't play a significant role in understanding why we are in this particular  vacuum amongst the vast number present in the landscape. Of course, we know that anti de Sitter vacua crunch on a time scale inversely proportional to their curvature. For de Sitter vacua, what really matters is the fact that the decay rate slows down exponentially  quickly as we descend through the de Sitter part of  landscape, until it comes to a halt for a Minkowski vacuum. In practice, this means de Sitter vacua of least curvature are the longest lived.

This mechanism takes us on a quantum journey through the landscape, leading us towards the vacuum of least absolute curvature. In a continuous landscape of vacua where a Minkowski vacuum is guaranteed, we are led to that vacuum, but in a discrete but dense landscape, we are led to the vacuum closest to Minkowski. The task of the model builder is to find a landscape where that near Minkowski vacuum has a curvature given by the current Hubble scale. This is achieved in the Bousso-Polchinski model (and its DBI extension) with relatively mild tunings of fundamental parameters and including of order a hundred species.

This suggests a new perspective on the cosmological constant problem as an alternative to the standard anthropic ideas usually employed in the presence of a landscape of vacua.  However, although we can explain how we were driven towards a  vacuum of low absolute curvature, we cannot explain how it is so young, at least in cosmological terms. Indeed, if the low curvature vacuum is so stable and long-lived, and matter is diluted over time,  how come we find ourselves in an era where the density of matter is comparable to that of the constant vacuum energy? This is the so-called coincidence problem  \cite{stein,Zlatev}. This is usually addressed with some new late time physics. For example, there could be a mechanism that forces the universe to end not long after dark energy begins to dominate, as was proposed in  \cite{phantom,lin,seq,Cunillera:2021izz}.

As a solution to the cosmological constant problem, the halting mechanism for membrane nucleation is undoubtedly interesting and one that can be incorporated into a wide class of models with  multiple three-form fields coupled to scalars.  Although this is often in tension with the WGC, that is not exclusively the case: there are elegant models such as the generalised DBI set up, where the halting mechanism exists for a family of membranes satisfying the WGC.  Nevertheless,  there are still some interesting and important problems that must be addressed, not least how it is aligned with a solution to the {\it why now?} problem.

\paragraph{Acknowledgements} 
AP was supported by STFC consolidated grant number ST/T000732/1 and YL by an STFC studentship. We would like to thank Jaume Garriga and Paul Saffin for useful discussions. 
For the purpose of open access, the authors have applied a CC BY public copyright licence to any Author Accepted Manuscript version arising. No new data were created during this study.

\appendix

\end{document}